\documentclass[12pt]{article}
\usepackage{amsmath,jstyle,amssymb,dsfont}

\newcommand{\arXiv}[1]{\href{http://www.arXiv.org/abs/#1}{arXiv:#1}}
\newcommand{\beq}{\begin{equation}}
\newcommand{\eeq}{\end{equation}}
\newcommand{\del}{\partial}

\makeatletter
\newcommand*\lslash
  {\text{\rlap{\kern.02em\@xxxii}}l}
\newcommand*\Lslash
  {\text{\rlap{\kern.03em\textbf{\@xxxii}}}L}
\makeatother

\def\d{\delta}
\def\i{\iota}
\def\p{\partial}
\def\s{\star}
\def\w{\wedge}
\def\pa{\partial}

\allowdisplaybreaks[4]

\author{\fontsize{13}{30}Sukruti Bansal$\,^{a}$\footnote{bansal.sukruti@gmail.com}\,, Oleg Evnin$\,^{a,b}\,$\footnote{oleg.evnin@gmail.com}\,, Karapet Mkrtchyan$\,^{c,d}\,$\footnote{k.mkrtchyan@imperial.ac.uk}}

\affiliation{$^a$Department of Physics, Faculty of Science, Chulalongkorn University,\\ \rule{1.8mm}{0mm}Phayathai Rd.,
Bangkok 10330, Thailand\vspace{1mm}\\
$^b$Theoretische Natuurkunde, Vrije Universiteit Brussel and\\ \rule{1.3mm}{0mm}The International Solvay Institutes, Pleinlaan 2, Brussels 1050, Belgium\vspace{1mm}\\
$^c$Scuola Normale Superiore and INFN, Piazza dei Cavalieri 7, 56126 Pisa, Italy\vspace{1mm}\\
$^d$Blackett Laboratory, Imperial College London SW7 2AZ, U.K.}

\title{\centering 
Polynomial Duality-Symmetric Lagrangians for Free $p$-Forms}

\abstract{We explore the properties of polynomial Lagrangians for chiral $p$-forms previously proposed by the last named author, and in particular, provide a self-contained treatment of the symmetries and equations of motion that shows a great economy and simplicity of this formalism.
We further use analogous techniques to construct polynomial democratic Lagrangians for general $p$-forms where electric and magnetic potentials appear on equal footing as explicit dynamical variables. Due to our reliance on the differential form notation, the construction is compact and universally valid for forms of all ranks, in any number of dimensions.
\vskip0.5cm

}

\begin{document}

\maketitle

\vspace{5in}

\section{Introduction}

Duality between electric and magnetic degrees of freedom has been a source of inspiration to theoretical physics since the formulation of Maxwell's theory of electromagnetism. In particular, it has prompted Dirac's influential treatment of magnetic charges. More contemporary related developments include the monopole condensation picture of confinement, Montonen-Olive duality, and various aspects of differential form fields ubiquitous in supersymmetric field theories.

While the electric-magnetic symmetry is apparent, say, in the vacuum Maxwell equations, creating a Lagrangian description of such duality-symmetric theories that would feature
electric and magnetic degrees of freedom on an equal footing is known to be involved. A closely related question is giving a Lagrangian description to chiral forms, invariant under electric-magnetic dualities and often arising in supersymmetric field theories. We provide a selection of relevant literature  \cite{Zw,DT,MSch,Sieg,KMkrtch,FJ,HT,BS,Harada,Tse1,MYW,Wot,Tse2,SchS,KhP,PST1,PST2,PSTproc,Tse3,DH,PSch,PST3,S5M,CW,BBS,MPS,PSTnotoph,RTse,MMMK,LM,KT,MMMK2,Sorokin,PST4,BH,KSV,Hiroshi,Sen1,AEShJ,Sen2,Buratti:2019cbm,Buratti:2019guq,Kiral,Lambert:2019diy,Townsend1,Andriolo:2020ykk,Townsend2,BLST,Bertrand:2020nob, Kosyakov,Vanichchapongjaroen,Bandos:2020hgy,Cremonini:2020skt}
that represents some of the history of the subject.

To set the stage, one can visualize the familiar example of the Maxwell equations in vacuum, which can be given using two-form field strength $F$ as
\begin{align}
d F=0\,,\quad d \star F=0\,. \label{Maxwell}
\end{align}
Solving the first equation using the Poincaré lemma, we introduce the vector potential: $F=dA$. Then, the second equation is a gauge-invariant wave equation for the vector $A$. We could also solve the second equation using Poincaré lemma, thus introducing a dual potential: $F=\star dB$ (in $d$ dimensions, $B$ is a $(d-3)$-form). In this case, the first equation of \eqref{Maxwell} is a non-trivial wave equation for the potential $B$. In both cases, it is straightforward to write a standard action in terms of potentials $A$ or $B$, which produce the corresponding equations.
The choice of one or another potential for the description of the theory breaks, however, its manifest electric-magnetic duality symmetry. Instead, one could aim at maintaining the duality manifest and keeping both potentials $A$ and $B$. In this case, \eqref{Maxwell} is equivalently re-encoded into the twisted selfduality relation
\begin{align}
d A = \star\, d B\,.\label{TD}
\end{align}
(The term `twist' denotes here the interchange of the two potentials that accompanies the Hodge dualization.) Lagrangian formulations giving rise to equation \eqref{TD} are much less obvious to construct than the corresponding formulations with only one of the two potentials.
Analogously and more generally, in $d$ space-time dimensions, equations of motion for a  free $p$-form can be written in terms of the twisted selfduality condition (\ref{TD}) employing both the curvature of the $p$-form potential, and its dual $(d-2-p)$-form potential.

Similarly, in the special cases of (anti)selfdual (chiral) fields, the equations can be given as duality relations
\begin{align}
d A=\pm \star d A\,. \label{D}
\end{align}
While in the cases of twisted selfduality equations \eqref{TD} it is possible to describe the corresponding degrees of freedom using two-derivative equations and straighforward (Maxwell) Lagrangians, the chiral fields force us to look for Lagrangians that give rise to first derivative equations \eqref{D}. There are several reasons why such Lagrangians are hard to construct, one of which is that the $(p+1)$-form equation \eqref{TD} cannot be generated as a variation of a scalar Lagrangian with respect to the $p$-form field $A$. This implies a necessity to break manifest Lorentz symmetry or introduce auxiliary fields.

A number of approaches to constructing Lagrangians for chiral forms and democratic formulations with both electric and magnetic potentials 
have appeared in the literature over the years, and it is beyond the limits of this introduction to critically review all of them; 
instead, we refer the reader to the original literature in our selection of references 
\cite{Zw,DT,MSch,Sieg,KMkrtch,FJ,HT,BS,Harada,Tse1,MYW,Wot,Tse2,SchS,KhP,PST1,PST2,PSTproc,Tse3,DH,PSch,PST3,S5M,CW,BBS,MPS,PSTnotoph,RTse,MMMK,LM,KT,MMMK2,Sorokin,PST4,BH,KSV,Hiroshi,Sen1,AEShJ,Sen2,Buratti:2019cbm,Buratti:2019guq,Kiral,Lambert:2019diy,Townsend1,Andriolo:2020ykk,Townsend2,BLST,Bertrand:2020nob, Kosyakov,Vanichchapongjaroen,Bandos:2020hgy,Cremonini:2020skt} and beyond. Our focus here will be on analyzing and developing a particular proposal put forth in \cite{Kiral}. This proposal is, in turn, rooted in the attractive approach originally developed in \cite{PST1,PST2,PST3} by Pasti, Sorokin and Tonin (PST). While the formalism of \cite{Kiral} arises from the PST theory
through introduction of an extra auxiliary form field via a particular version of the Hubbard-Stratonovich trick, it runs counter to the long perceived tension
between keeping Lagrangian descriptions of chiral forms local, manifestly Lorentz-invariant, ghost-free and polynomial with a finite number of terms and fields in the Lagrangian. Considerations of \cite{Kiral} have produced a theory that simultaneously possesses all of these appealing features.

The analysis of \cite{Kiral} has focused on designing a polynomial Lagrangian for chiral forms satisfying (\ref{D}), and showing its equivalence
to PST theory. In our present treatment, we start by taking up the result of the considerations of \cite{Kiral} and showing how to analyze the 
symmetries and equations of motion in this formalism. The result is, of course, the same as what one would get in PST theory (to which the formulation of \cite{Kiral} can be related), but it is visible from our treatment that the derivations are considerably more compact and transparent than what one would have had to undertake by reverting to PST theory first. This shows the efficiency of the polynomial formulation of \cite{Kiral}. To take the formulation
of \cite{Kiral} further, we then turn to the topic of democratic electric-magnetic theories that give a Lagrangian description to the duality relation (\ref{TD}) rather than the selfduality relation (\ref{D}). A formalism analogous to \cite{Kiral} can be developed for this objective, with a similarly high level of efficiency. This generates a compact treatment of democtratic polynomial Lagrangians for form fields of all ranks, in any number of dimensions.

%%%%%%%%%%%%%%%%%%%%%%%%%%%%%%%%%%%%%%

\section{Chiral $p$-forms}\label{selfdual}

We start by revisiting the polynomial formulations for a free chiral form proposed in \cite{Kiral}. Our purpose is twofold. First, we would like to give a demonstration of the algebraic efficiency of this formalism, which allows for an economical handling of the equations of motion and gauge symmetries. Second, we would like to thoroughly reformulate the formalism in the differential form notation, which makes it possible to treat all numbers of dimensions and form ranks in a uniform fashion. Much of the classic literature on related subjects \cite{PST1,PST2,PST3} relies on the index notation, where numerical coefficients of combinatorial nature appear in the formulas. The form notation eliminates such coefficients and makes the formulas more compact. (Some earlier treatments of related subjects relying on the form notation can be found in \cite{PST4,Hiroshi}.) We give conversion formulas connecting the form and index notation in appendix \ref{convDF}, and a collection of identities for differential forms in appendix \ref{excalid} (this material is completely standard, but it is convenient to have it handy while going through our subsequent derivations). Of these, the projection rejection identity
\beq
v\w \i_v A + \i_v (v\w A)=A,
\label{prre}
\eeq
valid for any unit vector $v$ and any form $A$, and the action of the Hodge star on the interior and exterior product,
\beq
\s \i_v (v\w A)=v\w\i_v(\s A),\qquad \s (v\w i_vA)=\i_v(v\w\s A),
\label{starivv}
\eeq
will be encountered especially frequently in our derivations (normally, with $da/\sqrt{(\del a)^2}$ playing the role of $v$ for a scalar field $a$). In a slight abuse of notation, we shall employ the same letter to refer to vectors and their dual 1-forms, but there should be no situations where this could cause confusion. We shall also use the notation
\beq
A^2\equiv A\w\s A.
\eeq

The formulation of \cite{Kiral} comes in two versions, related by a simple field redefinition, but endowed with rather different flavors. In the first formulation introduced in \cite{Kiral}, the field content is a gauge $p$-form, an auxiliary $p$-form satisfying an algebraic equation of motion, and an auxiliary scalar field playing the same role as in PST formalism. The advantages of this formalism are its immediate relation to PST theory (which is recovered upon integrating out the auxiliary $p$-form) and the simpler form of some of its gauge symmetries. The second formulation appears, at the first sight, to have a doubled field content consisting of two $p$-form gauge fields and the same auxiliary scalar as before. The truth is that, because of the rich gauge symmetry, the actual dynamical content of the theory is halved rather than doubled, leaving a single chiral $p$-form. The advantage of this formalism is its very conventional appearance in terms of two ordinary $p$-forms with Maxwellian gauge symmetries.

In what follows, we shall first review the `Maxwellian' formalism and provide a complete self-contained analysis of the symmetries and the equations of motion, showing the algebraic appeal of this formulation. Then for the convenience of the reader we will summarize the structure of the formalism with an algebraic auxiliary field.

%%%%%%%%%%%%%%%%%%%%
 
\subsection{Formulation with a doubled set of Maxwellian fields}

Following \cite{Kiral}, we consider the Lagrangian
\begin{align}\label{FQlag}
    \mathcal{L}&=(F+a\,Q)^2+2\,aF\w Q\,,
\end{align}
where $F=dA$ and $Q=dR$. Here, $A$ and $R$ are $p$-forms and $a$ is a scalar field. A $p$-form may have a selfdual field strength in $d=2p+2$ dimensions, and furthermore, in Minkowski signature, $p$ must be even, which is what we shall assume until the end of section \ref{selfdual}. (Note that, in this situation, $\s\s=+1$ when acting on any $(p+1)$-form, which we shall apply automatically throughout until the end of section \ref{selfdual}.) Note the very conventional field content with two Maxwellian $p$-form gauge fields $A$ and $R$, only entering the Lagrangian through their respective field strengths $F$ and $Q$. 

Lagrangian \eqref{FQlag} possesses the following four gauge symmetries:
\begin{align}
    \d a&=0\,, \,\, \d A= dU\,, \,\, \d R=0\,; \label{AsymmKar211} \\
    \d a&=0\,, \,\, \d A=0 , \,\, \d R=dU\,; \label{RsymmKar211} \\
    \d a&=0, \,\, \d A=-\,a\,da\w U\,,  \,\, \d R=da\w U\,; \label{ARgaugesym}\\
\d a&=\varphi\,, \,\, \d A=-\,\frac{a\,\varphi}{(\pa a)^2}\,\i_{da}(Q+\s Q)\,, \,\, \d R=\frac{\varphi}{(\pa a)^2}\,\i_{da}(Q+\s Q)\,,\label{PST3FQ}
\end{align}
where $\varphi$ is an arbitrary scalar function and $U$ is an arbitrary $(p-1)$-form (independently specified for each of the three transformations where it is featured). The first two transformations are Maxwellian shifts of $A$ and $R$, obvious as the Lagrangian only depends on the corresponding field strengths. The third symmetry leaves the first term of (\ref{FQlag}) unchanged and modifies the  second one by a total derivative, which vanishes upon integration. This symmetry will be crucial below for establishing the selfduality of the physical degrees of freedom. Validating the last transformation requires slightly more work. The variation of Lagrangian \eqref{FQlag} under arbitrary shifts of $a$, $A$ and $R$, up to total derivative terms, is
\begin{align}
    \d\mathcal{L}=&\,2\,\big[\d a\{(F+a\,Q)\w\s Q+F\w Q\}-\d A\w[d\,\{\s(F+aQ)\}+da\w Q] \nonumber \\
    &-\d R\w[d\,\{a\s(F+aQ)\}-da\w F]\big]\label{FQlagrvar}
\end{align}
Note that the content of the square brackets  in the last line can be recast as~ $a\,d[\s(F+aQ)]+da\w [\s(F+aQ)-F]$. Thereafter, assuming $\d A=-\,a\,\d R$, which holds for \eqref{PST3FQ}, one gets
\beq
	\d\mathcal{L}=2\,[\d a\{(F+aQ)\w\s Q+F\w Q\}+\d R\w da\w\{(F+aQ)-\s(F+aQ)\}],
\label{dAisadR}
\eeq
or
\begin{align}\label{FQPSTver}
    \d\mathcal{L}=2\,\Big[\varphi\,(H\w\s Q+F\w Q)+\frac{\varphi}{(\pa a)^2}\,da\w\i_{da}(Q+\s Q)\w(H-\s H)\Big],
\end{align}
where we have introduced $H\equiv F+aQ$ for brevity. The second $\varphi$-term above gets simplified as follows:
\begin{align}
    \frac{\varphi}{(\pa a)^2}&\,da\w\i_{da}(Q+\s Q)\w(H-\s H) \nonumber \\
 &=\frac{\varphi}{(\pa a)^2}[(\pa a)^2(Q+\s Q)-\i_{da}\{da\w(Q+\s Q)\}]\w(H-\s H) \quad \text{(by \eqref{prre})} \nonumber\\
    &=\big[\varphi\,(Q+\s Q)-\frac{\varphi}{(\pa a)^2}\s\{da\w\i_{da}(Q+\s Q)\}\big]\w(H-\s H) \quad \text{(by \eqref{starivv})} \nonumber\\
    &=-\,2\,\varphi\,(H\w\s Q+F\w Q)-\frac{\varphi}{(\pa a)^2}\s(H-\s H)\w\{da\w\i_{da}(Q+\s Q)\} \quad \text{(by \eqref{AstarC})} \nonumber\\
    &=-\,2\,\varphi\,(H\w\s Q+F\w Q)-\frac{\varphi}{(\pa a)^2}\,da\w\i_{da}(Q+\s Q)\w(H-\s H). \nonumber
\end{align}
Taking the second term to the left-hand side, one concludes that
\beq
\frac{\varphi}{(\pa a)^2}\,da\w\i_{da}(Q+\s Q)\w(H-\s H)=-\,\varphi\,(H\w\s Q+F\w Q).
\eeq
Substituting this formula back into \eqref{FQPSTver} results in $\d\mathcal{L}=0$, and hence (\ref{PST3FQ}) is a valid symmetry.

Note that transformations (\ref{PST3FQ}) imply that the field $a$ can be shifted arbitrarily, and is thus a pure gauge degree of freedom. Admissible gauges, as in PST formalism \cite{PST3}, are those that provide $a$ as a good global coordinate function on the Minkowski space. In particular, gradients of $a$ should not vanish anywhere, and the level hypersurfaces $a=\mathrm{const}$ must be of topology $R^{d-1}$ and furnish a globally nondegenerate foliation of $R^d$. This, obviously, implies that one cannot choose $a$ to be a constant function (in this degenerate case, the formalism produces a single nonchiral propagating $p$-form, instead of a chiral $p$-form).

Among all the commutators of the symmetry transformations \eqref{AsymmKar211} to \eqref{PST3FQ}, the only non-zero commutators are those of transformation \eqref{PST3FQ} with itself and \eqref{ARgaugesym}. The commutator of \eqref{ARgaugesym} with \eqref{PST3FQ} is
\begin{align}
    \d_U\d_\varphi-\d_\varphi\d_U=\d_V+\d_W+\d_X,
\end{align}
where $\d_V$ is transformation \eqref{AsymmKar211} with $V=a\varphi U$ substituted for $U$, $\d_W$ is transformation \eqref{RsymmKar211} with $W=-\varphi U$ substituted for $U$, and $\d_X$ is transformation \eqref{ARgaugesym} with $X=(\varphi\,\i_{da}dU)/(\pa a)^2$ substituted for $U$. The commutator of tranformation \eqref{PST3FQ} with parameters $\varphi_1$ and $\varphi_2$ is 
\begin{align}
    \d_{\varphi_1}\d_{\varphi_2}-\d_{\varphi_2}\d_{\varphi_1}=\d_V,
\end{align} 
where $\d_V$ is transformation \eqref{ARgaugesym} with $V=(\varphi_1\,\i_{d\varphi_2}-\varphi_2\,\i_{d\varphi_1})\i_{da}(Q+\s Q)$ substituted for $U$.

From (\ref{FQlagrvar}), the equations of motion are
\begin{align}
    E_a\equiv\frac{\d\mathcal{L}}{\d a}&\equiv (F+aQ)\w\s Q+F\w Q=0\,,\label{EawithY} \\
    E_A\equiv\frac{\d\mathcal{L}}{\d A}&\equiv d\,[\s(F+aQ)]+da\w Q=0\,, \label{EAwithY} \\
    E_R\equiv\frac{\d\mathcal{L}}{\d R}&\equiv d\,[a\s(F+aQ)]-da\w F=0\,. \label{ERwithY}
\end{align}
We would like to examine the combination $E_R-aE_A$, which can be simply read off the $\d R$-term in (\ref{dAisadR}), or recovered directly from (\ref{EAwithY}-\ref{ERwithY}) in the following form by operations similar to the derivation of (\ref{dAisadR}):
\beq
da\w [F+aQ-\s(F+aQ)]=0.
\label{daFQ}
\eeq
Applying a Hodge star to the above gives $\i_{da}[F+aQ-\s(F+aQ)]=0$. Acting on this last equation with $da\w$, acting on (\ref{daFQ}) with $\i_{da}$, adding the results and applying the projection-rejection identity (\ref{prre}), one gets
\beq
F+aQ=\s(F+aQ).
\label{FQselfd}
\eeq 
In other words, $F+aQ$ is selfdual on any solutions of the equations of motion.

While (\ref{FQselfd}) has been derived from (\ref{EAwithY}-\ref{ERwithY}), it ensures that (\ref{EawithY}) is automatically satisfied. Indeed, for any two forms $A$ and $B$ of the same rank, $A\w\s B=B\w\s A$  (this combination is often called simply $A\s B$). Hence, the left-hand side of (\ref{EawithY}) is rewritten as $Q\w\s(F+aQ)+F\w Q=Q\w(F+aQ)+F\w Q$, which is identically zero by (\ref{commext}) since the ranks of both $F$ and $Q$ are odd. This lack of a constraining dynamical equation for $a$ is another manifestation of the gauge symmetry (\ref{PST3FQ}).

It remains to reveal that the full dynamical content of our theory is a single chiral $p$-form. To this end, we substitute (\ref{FQselfd}) into (\ref{EAwithY}) to obtain
\beq
da\w dR=0.
\eeq
Equations of this form are ubiquitous within PST theory and its relatives, and appear, in particular, among the PST equations of motion \cite{PST3,Hiroshi}. We give a careful derivation of the general solution in appendix \ref{appgensol}. In brief, this equation is solved by restricting it to hypersurfaces\footnote{By $a(x)=\mathrm{const}$, we of course mean hypersurfaces defined by a specific value of a non-constant function $a(x)$, and not that $a(x)$ is a constant function, which would have been an inadmissible gauge.} $a=\mathrm{const}$, where it says simply that a tangential restriction of $R$ to each such hypersurface is a closed form within the hypersurface. In this form, the equations are immediately integrated to yield
\beq
R=dU+da\w V,\label{R}
\eeq
with arbitrary $U$ and $V$.
The first term can be completely gauged away by (\ref{RsymmKar211}), while the second term can be completely gauged away by (\ref{ARgaugesym}). As a result, the most general solution can be gauged to
\beq
R=0,
\eeq
whereupon (\ref{FQselfd}) gives
\beq
F=\s F.
\label{selfF}
\eeq
We have thus proved that for any solution to (\ref{EawithY}-\ref{ERwithY}), $a$ can be changed arbitrarily and $R$ can be gauged to 0 by the symmetries of the theory, and then the field strength of $A$ becomes selfdual. Hence, the full dynamical content of (\ref{FQlag}) is a single propagating chiral $p$-form.

\subsection{Formulation with an algebraic auxiliary field}

We now recast Lagrangian (\ref{FQlag}) into an alternative form such that no derivatives of $R$ appear in the Lagrangian. Redefining the field $A$ as $A_\mathrm{old}=A_\mathrm{new}-aR$, we get
\beq\label{FRlag}
    \mathcal{L}=(F-da\w R)^2+2\,da\w F\w R.
\eeq
The above Lagrangian has the following four gauge symmetries, which are direct counterparts of (\ref{AsymmKar211}-\ref{PST3FQ}):
\begin{align}
    \d a&=0\,, \quad \d A= dU\,, \quad \d R=0; \label{ARsymmMax}\\
    \d a&=0\,, \quad \d A=0 , \quad \d R=da\w U\,; \label{RsymmKarlag} \\
    \d a&=0\,, \quad \d A= -\,da\w U\,, \quad \d R=dU; \label{ARsymmKarlag}\\
    \d a&=\varphi\,, \quad \d A=\varphi R\,, \quad \d R=\frac{\varphi}{(\pa a)^2}\,\i_{da}(dR+\s dR)\,. \label{PST3}
\end{align}
While these symmetries are guaranteed to work due to the straightforward relation 
between (\ref{FRlag}) and (\ref{FQlag}), we also demonstrate how to validate them directly
for the convenience of the reader.
Verification of the first three symmetries is straightforward: (\ref{ARsymmMax}) does not change (\ref{FRlag}) as the latter only depends on $F$;
(\ref{RsymmKarlag}) does not change (\ref{FRlag}) as $R$ only enters through $da\w R$; (\ref{ARsymmKarlag}) leaves the first term in (\ref{FRlag}) invariant, while the variation of the second term is a total derivative that vanishes upon integration. For (\ref{PST3}), the first term of  (\ref{FRlag}) transforms as
\begin{align}
&\d [(F-da\w R)^2] \nonumber\\
&\quad=2\,(F-da\w R)\w\s\Big[\varphi\,dR-\frac{\varphi}{(\pa a)^2}\,da\w\i_{da}(dR+\s dR)\Big] \nonumber\\
&\quad=\frac{2\,\varphi}{(\pa a)^2}\,(F-da\w R)\w[\s \i_{da}(da\w dR)-\i_{da}(da\w dR)]\nonumber\\
&\quad=\frac{2\,\varphi}{(\pa a)^2}\,\Big[F\w\{\s \i_{da}(da\w dR)-\i_{da}(da\w dR)\}+da\w R\w\i_{da}(da\w dR)\Big]\nonumber\\
&\quad=\frac{2\,\varphi}{(\pa a)^2}\,F\w[\s \i_{da}(da\w dR)-\i_{da}(da\w dR)]+2\,\varphi\, da\w R\w dR,\nonumber
\end{align}
and second term transforms as
\begin{align}
2\,\d(F\w da\w R)&=2\,d\varphi\w R\w da\w R+2\,\varphi\,dR\w da\w R  \nonumber\\
&\hspace{2cm}+ 2\,F\w d\varphi\w R+2\,F\w da\w\frac{\varphi}{(\pa a)^2}\,\i_{da}(dR+\s dR) \nonumber\\
&=2\,d(\varphi\,da\w R\w R)+2\,\varphi\,da\w R\w dR-2\,d(\varphi\,F\w R)\nonumber\\
&\hspace{2cm}-2\,\varphi\,F\w dR +2\,F\w da\w\frac{\varphi}{(\pa a)^2}\i_{da}(dR+\s dR)\nonumber\\
&=2\,d(\varphi\,da\w R\w R-\varphi\,F\w R)+2\,\varphi\,da\w R\w dR \nonumber\\
&\hspace{2cm}+\frac{2\,\varphi}{(\pa a)^2}\,F\w [(-\,\i_{da}(da\w dR)+\s\i_{da}(da\w  dR)]. \nonumber
\end{align}
Subtracting the second variation from the first leaves a total derivative, which yields zero upon integration.

As mentioned above, and discussed in detail in \cite{Kiral}, solving the (algebraic) equation of motion for $R$ and substituting the result into (\ref{FRlag}) recovers the (nonpolynomial) PST Lagrangian. As a consequence, the standard analysis of PST formalism \cite{PST3, Hiroshi} guarantees that the dynamical content of (\ref{FRlag}) amounts to a single chiral $p$-form, and no other propagating degrees of freedom.
That is, of course, in agreement with what we have already shown for Lagrangian (\ref{FQlag}) entirely within the context of the polynomial theory.

%%%%%%%%%%%%%%%%%%%%%%%%%%%%%%%%%%%%%%%%%%%%

\section{Democratic polynomial duality-symmetric Lagrangians}

We now turn to an ordinary nonchiral Maxwellian $p$-form, which could of course be described with the Lagrangian $F^2$. However, we would like to represent it in a formalism where the electric potential $A$ and the magnetic potential $B$ both appear as explicit dynamical variables, while the corresponding field strengths $F=dA$ and $G=dB$ satisfy the electric-magnetic duality relation
\beq
F=\s G.
\label{FG}
\eeq
This gives the problem a structure rather similar to the chiral form story of the previous section.

Democratic formulations of this type, where electric and magnetic degrees of freedom appear on equal footing are of interest in the context of supersymmetric theories, as well as in cases where it is necessary
to couple the $p$-form to both electric and magnetic charges, see \cite{LM}. Such theories have previously
been considered in \cite{PST1,PST2,PST4} using the nonpolynomial PST formalism, and in specific numbers
of dimensions. Our goal is to construct a polynomial formulation akin to the chiral forms of section \ref{selfdual}, relying on the differential form notation and keeping the treatment applicable to differential forms of any
rank in any number of dimensions.

As in section \ref{selfdual}, there are two closely related versions of our formalism. One can either have a doubled set of Maxwellian forms and an extra auxiliary scalar that gets reduced to a single $p$-form dynamically due to the gauge symmetries, or instead of the doubled set of Maxwellian forms, one can have electric and magnetic potentials and two additional auxiliary form fields satisfying algebraic equations of motion.
This latter formalism is more directly related to PST-like nonpolynomial formulation, which is obtained by
integrating out the two algebraic auxiliary fields.

\subsection{Formulation with a doubled set of Maxwellian fields}

We propose the following Lagrangian, which is overall a direct generalization of (\ref{FQlag}) adapted to condition (\ref{FG}) rather than (\ref{selfF}), though it is essential to fix some signs and the ordering of forms in wedge-products:
\begin{align}\label{democ}
    \mathcal{L}&=(F+aP)^2+(G+a\,Q)^2-2\,a\,Q\w F+2\,a\,G\w P. 
\end{align}
We are in $d$ spacetime dimensions; $a$ is a scalar field; $F=dA$ where $A$ is a $p$-form gauge potential (which can be thought of as electric); $G=dB$ where $B$ is a $(d-p-2)$-form gauge potential (which can be thought of as magnetic); $P=dR$ where $R$ is a $p$-form field; $Q=dS$ where $S$ is a $(d-p-2)$-form field, with these last two forms playing an auxiliary role. The values of $p$ and $d$ are completely arbitrary in this section. Note that $\s\s F= (-1)^{(p+d+pd)}F$ according to (\ref{starstarid}), while $F\w Q=-(-1)^{(p+d+pd)}Q\w F$ according to (\ref{commext}). 

The variation of the Lagrangian is given by:
\begin{align}
    \d\mathcal{L}=-\,2\,\Big[&\d a\,[\{(-1)^{(p+d+pd)}\s(F+aP)-G\}\w P-Q\w\{\s(G+aQ)-F\}] \nonumber\\
    &+(-1)^p\,\d A\w d[\s(F+aP)+(-1)^{(p+d+pd)}aQ]\nonumber\\
    &+(-1)^{(d-p)}\,\d B\w d[\s(G+a\,Q)+aP] \nonumber\\
    & +(-1)^p\,\d R\w d[a\s(F+aP)-(-1)^{(p+d+pd)}a\,G] \nonumber\\
    &+(-1)^{(p+d)}\,\d S\w d[a\s(G+a\,Q)-aF]\Big]\label{varlagrFPGQ}
\end{align}
The theory is invariant under the following gauge transformations:
\begin{align}
    \d a&=0\,, \,\, \d A= dU\,, \,\, \d B=0\,,\,\, \d R=0\,,\,\, \d S=0\,; \label{maxA}\\
    \d a&=0\,, \,\, \d A=0\,, \,\, \d B=dV\,,\,\, \d R=0\,,\,\, \d S=0\,; \label{maxB}\\
    \d a&=0\,, \,\, \d A=0 , \,\, \d B=0\,,\,\, \d R=dU\,,\,\, \d S=0\,; \label{maxR}\\
    \d a&=0\,, \,\, \d A=0 , \,\, \d B=0\,,\,\, \d R=0\,,\,\, \d S=dV\,; \label{maxS}\\
    \d a&=0\,, \,\, \d A=-\,a\,da\w U\,, \,\, \d B=0\,,  \,\, \d R=da\w U\,,\,\, \d S=0\,; \label{daR}\\
    \d a&=0\,,\,\, \d A=0\,,\,\,\d B=-\,a\,da\w V\,,\,\,\d R=0\,,\,\, \d S=da\w V\,; \label{daS} \\
    \d a&=\varphi\,, \,\, \d A=-\,\frac{a\,\varphi}{(\pa a)^2}\,\i_{da}[P+(-1)^{(p+d+pd)}\s Q]\,,\,\, \d B=-\,\frac{a\,\varphi}{(\pa a)^2}\,\i_{da}(\s P+Q)\,, \nonumber\\
    \d R&=\frac{\varphi}{(\pa a)^2}\,\i_{da}[P+(-1)^{(p+d+pd)}\s Q]\,,\,\,
    \d S=\frac{\varphi}{(\pa a)^2}\,\i_{da}(\s P+Q)\,; \label{PST3PQ}
\end{align}
where $\varphi$ is an arbitrary scalar, $U$ is an arbitrary $(p-1)$-form and $V$ is an arbitrary $(d-p-3)$-form (the latter two forms specified completely independently in the different transformations where they appear). The symmetries can be verified by methods analogous to the previous section.

All the above transformations commute except the following three. The commutator of transformation \eqref{daR} with \eqref{PST3PQ} is
\begin{align}
    \d_U\d_\varphi-\d_\varphi\d_U=\d_V+\d_W+\d_X,
\end{align}
where $\d_V$ is transformation \eqref{maxA} with $V=a\varphi U$ substituted for $U$, $\d_W$ is transformation \eqref{maxR} with $W=-\varphi U$ substituted for $U$, and $\d_X$ is transformation \eqref{daR} with $X=(\varphi\,\i_{da}dU)/(\pa a)^2$ substituted for $U$. There is a directly analogous expression for the commutator of  \eqref{daS} with \eqref{PST3PQ}.  The commutator of two transformations given by \eqref{PST3PQ} with parameters $\varphi_1$ and $\varphi_2$ is 
\begin{align}
    \d_{\varphi_1}\d_{\varphi_2}-\d_{\varphi_2}\d_{\varphi_1}=\d_X+\d_Y,
\end{align}
where $\d_X$ is transformation \eqref{daR} with $X=(\varphi_1\,\i_{d\varphi_2}-\varphi_2\,\i_{d\varphi_1})\i_{da}[P+(-1)^{(p+d+pd})\s Q]$ substituted for $U$ and $\d_Y$ is transformation \eqref{daS} with $Y=(\varphi_1\,\i_{d\varphi_2}-\varphi_2\,\i_{d\varphi_1})\i_{da}(\s P+Q)$ substituted for $V$.

The equations of motion can be extracted from (\ref{varlagrFPGQ}) as
\begin{align}
    &E_a\equiv [(-1)^{(p+d+pd)}\s(F+aP)-G]\w P-Q\w[\s(G+aQ)-F]=0, \label{eoma}\\
    &E_A\equiv d[\s (F+aP)]+(-1)^{(p+d+pd)}da\w Q=0, \label{eomA}\\
    &E_B\equiv d[\s(G+a\,Q)]+da\w P=0 ,\label{eomB}\\
    &E_R\equiv d[a\s(F+aP)]-(-1)^{(p+d+pd)}da\w G=0, \label{eomR}\\
    &E_S\equiv d[a\s(G+a\,Q)]-da\w F=0. \label{eomS}
\end{align}
We then proceed, much as under (\ref{ERwithY}), by forming the combination
\begin{align}
	E_R-aE_A\equiv da\w [\s(F+aP)-(-1)^{(p+d+pd)}(G+aQ)]=0.
\end{align}
Taking Hodge dual of the above and then using identity \eqref{ivstarid}, we get
\begin{align}
	\i_{da}[(F+aP)-\s(G+aQ)]=0.
\label{idaFP}
\end{align}
We also have
\begin{align}
	E_S-aE_B\equiv da\w[\s(G+aQ)-(F+aP)]=0.
\label{daFP}
\end{align}
Acting with $da\w$ on (\ref{idaFP}), acting with $\i_{da}$ on (\ref{daFP}), adding the results and using the projection-rejection identity (\ref{prre}), we get
\begin{align}\label{FaPstarGaQ}
	(F+aP)=\s(G+aQ)
\end{align}
As in the previous section, this relation, derived from the equations of motion (\ref{eomA}-\ref{eomS}), ensures that the equation of motion for $a$ given by  \eqref{eoma} is identically satisfied. 

We now proceed to reveal the dynamical content of (\ref{democ}).
Plugging the expression for $\s(F+aP)$ derived from the above relation into \eqref{eomA}, and plugging the above expression for $\s(G+aQ)$ into \eqref{eomB} we get
\begin{align}
	da\w dS=0 \quad \text{and} \quad da\w dR=0
\end{align}
respectively.
The general solution of these equations is obtained as in appendix \ref{appgensol} and reads
\begin{align}
	R=dU+da\w V, \qquad S=dW+da\w X,
\end{align}
where $U$ and $V$ are arbitrary $p$-forms and $W$ and $X$ are arbitrary $(d-p-2)$-forms. In the above solution for $R$, $dU$ is gauged away by the Maxwellian shift symmetry \eqref{maxR} and $da\w V$ is gauged away using symmetry \eqref{daR}. And in the above solution for $S$, $dW$ is gauged away by transformation \eqref{maxS} and $da\w X$ is gauged away using symmetry \eqref{daS}. Then we get
\begin{align}
	R=S=0\,.
\end{align}
Plugging this solution for $R$ and $S$ into relation \eqref{FaPstarGaQ} we get
\begin{align}
	F=\s G
\end{align}
Thus, despite starting with a theory containing four form fields, the gauge symmetries reduce the physical degrees of freedom down to a single Maxwellian $p$-form, with electric and magnetic field strengths $F$ and $G$ respectively.

%%%%%%%%%%%%%%%%%%

\subsection{Formulation with algebraic auxiliary fields}

As for the selfdual case of section \ref{selfdual}, by applying 
the field redefinitions $A_\mathrm{old}= A_\mathrm{new}-aR$ and $B_\mathrm{old} = B_\mathrm{new}-aS$, one can recast (\ref{democ}) in a form where the derivatives of the auxiliary forms $R$ and $S$ do not appear in the Lagrangian:
\begin{align}\label{polylag}
    \mathcal{L}=(F-da\w R)^2+(G-da\w S)^2+2\, da\w S\w F-2\,G\w da\w R.
\end{align}
With respect to these variables, the gauge transformations take the following form:
\begin{align}
    \d a&=0\,, \,\, \d A= dU\,, \,\, \d B=0\,,\,\, \d R=0\,,\,\, \d S=0\,; \\
    \d a&=0\,, \,\, \d A= 0\,, \,\, \d B=dV\,,\,\, \d R=0\,,\,\, \d S=0\,; \\
    \d a&=0\,, \,\, \d A=0 , \,\, \d B=0\,,\,\, \d R=da\w U\,,\,\, \d S=0; \label{gaugeR}\\
    \d a&=0\,, \,\, \d A=0 , \,\, \d B=0\,,\,\, \d R=0\,,\,\, \d S=da\w V; \label{gaugeS}\\
    \d a&=0\,, \,\, \d A= da\w U\,,\,\, \d B=0\,,\,\, \d R=-\,dU\,,\,\, \d S=0\,; \\
    \d a&=0\,, \,\, \d A=0 \,,\,\, \d B=da\w V\,,\,\, \d R=0\,,\,\, \d S=-\,dV\,; \\
    \d a&=\varphi\,, \,\, \d A=\varphi R\,,\,\, \d B=\varphi S\,, \,\, \d R=\frac{\varphi}{(\pa a)^2}\,\i_{da}[dR+(-1)^{(p+d+pd)}\s dS]\,, \nonumber\\
    \d S&=\frac{\varphi}{(\pa a)^2}\,\i_{da}(\s dR+dS)\,.
\end{align}

While this form of the Lagrangian looks less conventional than (\ref{democ}) relative to textbook field theories, it has the advantage that the equations of motion for $R$ and $S$ are purely algebraic. One can then immediately solve these equations, substitute the result back into the Lagrangian, and thus obtain a
nonpolynomial PST-like formulation with only two forms $A$ and $B$ that connects to the considerations
of \cite{PST1,PST2,PST4,LM}. We give the relevant derivations in appendix \ref{appequiv}.

%%%%%%%%%%%%%%%%%%

\subsection{A simple example in $D=3$}

For clarity, we present the simplest example of a polynomial democratic Lagrangian consisting of a scalar ($p=0$) $\varphi$ and a vector field ($d-p-2=1$) $A_\m$ in three dimensions. This may be useful for readers more accustomed to the index notation.  (Conversion formulas for switching between form and index notation are given in appendix \ref{convDF}). The field strengths are
\begin{align}
F_\m=\partial_\m \varphi\,,\quad G_{\m\n}=2\,\partial_{[\m} A_{\n]}.
\end{align}
The polynomial Lagrangian is
\begin{align}\label{3dlag}
    \mathcal{L}&=F^2+G^2+(da\w R)^2+(da\w S)^2-2\,\mathcal{F}\w\s(da\w R)+2\,\s\mathcal{F}\w\s(da\w S) \nonumber\\
    &\equiv F_\m F^\m +\frac12\, G_{\m\n}G^{\m\n}+\pa_\m a R\,\pa^\m a R + \pa_{[\m}a\,S_{\n]}\pa^\m a\, S^\n-2\,\mathcal{F}_\m\pa^\m a R+\epsilon_{\r\m\n}\,\mathcal{F}^\r\,\pa^\m a\,S^\n
\end{align}
where $R$ is a rank-$0$ (scalar) auxiliary field, $S$ is a rank-$1$ (vector) auxiliary field and $\mathcal{F}_\m=F_\m+\frac12\, \epsilon_{\m\n\r}G^{\n\r}$.
The equation of motion for the field $R$ is
\beq\label{solR0}
   da\w\s(da\w R -\mathcal{F})=0,\quad \mathrm{or}\quad \pa_\m a\,\pa^\m a R- \pa_\m a \mathcal{F}^\m =0.
\eeq
This is solved, up to gauge transformations (\ref{gaugeR}), by
\beq
 R=\frac{\pa_\m a}{(\pa a)^2}\mathcal{F}^\m.
\eeq
Analogously for $S$,
\beq\label{solS1}
   S=-\,\frac{\i_v \mathcal{\s F}}{\sqrt{(\partial a)^2}}, \quad \mathrm{or}\quad
 S_\r=-\,\frac{1}{\sqrt{(\pa a)^2}}\,\epsilon_{\m\n\r}\,\mathcal{F}^\m v^\n\,.
\eeq
Substituting solutions \eqref{solR0} and \eqref{solS1} into Lagrangian \eqref{3dlag} we get,
\begin{align}
    \mathcal{L}&=F_\m F^\m +\frac12\, G_{\m\n}G^{\m\n}-v^\m\mathcal{F}_\m\mathcal{F}^\n v_\n-v^{[\m}\mathcal{F}^{\n]}\mathcal{F}_\n v_\m \nonumber \\
    &=F_\m F^\m +\frac12\, G_{\m\n}G^{\m\n}-v^\m\mathcal{F}_\m\mathcal{F}^\n v_\n-\frac12\,v^{\m}\mathcal{G}_{\m\n}\,\mathcal{G}^{\r\n} v_\r
\end{align}
where $\mathcal{G}_{\m\n}=\epsilon_{\m\n\r}\,\mathcal{F}^\r$.
The Lagrangian above is manifestly in a PST-like form.

\section{Discussion} \label{disc}

We have revisited the polynomial Lagrangian formulation for chiral $p$-forms proposed in \cite{Kiral} and demonstrated that it leads to a compact,
efficient treatment of selfduality. We have then extended this formulation to the case of polynomial democratic Lagrangians where the dual electric and magnetic potentials appear on the same footing. Due to our reliance on the differential form notation, the formulation is universally valid for forms of
all ranks in any number of dimensions. We conclude with a summary of related situations where analogous approaches could be of value. This summary 
focuses on classical field theories, though having polynomial formulations at hand could also be beneficial for quantization.

One motivation for developing simple and tractable theories of free fields is that it may aid a subsequent inclusion of interactions.
Interactions of chiral fields are strongly constrained \cite{PSch}, but a PST-like theory for DBI interactions of chiral forms is known \cite{S5M}.
It could be interesting to look for a polynomial version of this theory along the lines similar to our current treatment. Another interesting question
is the number of auxiliary PST scalars in the theory. Since the auxiliary PST scalar is required to have properties of a global coordinate, both in the original PST theory and in its polynomial version of \cite{Kiral}, a natural inclination is to add extra scalars of this kind on a many-dimensional manifold. One particular construction of this sort has been proposed in \cite{KSV}. It could be interesting to polynomialize it with the techniques of \cite{Kiral} and look for formulations with a still larger number of PST scalars compared to \cite{KSV}.

Turning to problems involving democratic formulations with explicit electric and magnetic potentials, a foundational question that goes back to
\cite{Zw} is developing field theories including both electric and magnetic matter. There are no known ways to give a consistent Lagrangian descriptions
when both electrically and magnetically charged classical fields are present, though the theories become consistent upon quantization, subject to the Dirac condition on electric and magnetic charges \cite{BS,LM}. This paradoxical situation would be worth revisiting. 

Finally, there has been a recent surge of interest \cite{Townsend1,Buratti:2019cbm,Buratti:2019guq,BLST,Kosyakov,Bandos:2020hgy} in interacting theories of p-form fields with electric-magnetic duality symmetries, see also earlier related works \cite{BB1,BB,Chruscinski}. It could be worthwhile to approach such theories from a perspective similar to our current treatment.

\section*{Acknowledgements}

We have benefitted from discussions with Hiroshi Isono, Akash Jain, Dmitri Sorokin and Arkady Tseytlin. Research of S.B. and O.E. is supported by the CUniverse research promotion initiative (CUAASC) at Chulalongkorn University. K.M. is supported by the European Union’s Horizon 2020 research and innovation programme under the Marie Sk\lslash odowska-Curie grant number 844265.

\appendix

\section{Conversion between differential calculus and index notation}\label{convDF}

In the following, the Levi-Civita tensor is understood as
\begin{align}
    \epsilon_{\m_1\m_2...\m_d}=\sqrt{|g|}\,\varepsilon_{\m_1\m_2...\m_d}\,, \qquad \epsilon^{\m_1\m_2...\m_d}=\frac1{\sqrt{|g|}}\,\varepsilon_{\m_1\m_2...\m_d}\,,
\label{LCtens}
\end{align}\\
given in terms of the metric deteminant $g=\text{det}(g_{\m\n})$ and the Levi-Civita symbol
\begin{align}\label{LeviCivitensdens}
    \varepsilon_{\m_1\m_2...\m_d}= \begin{cases} +1\,\, \text{if}\,\, (\m_1, \m_2, ..., \m_d)\,\, \text{is an even permutation of}\,\, (1, 2, ..., d)\\
    -1\,\, \text{if}\,\, (\m_1, \m_2, ..., \m_d)\,\, \text{is an odd permutation of}\,\, (1, 2, ..., d)  \\
    0\,\,\text{otherwise}\,.
    \end{cases}
\end{align}

\vspace{3mm}
\noindent 
\textbf{Component representation:}
\begin{align}
    &A^{(p)}=\frac{1}{p!}\,\left(A_{[\m_1\m_2...\m_p]}\right)\,dx^{\m_1}\w dx^{\m_2}\w...dx^{\m_p}\,.
\end{align}
\textbf{Exterior product:}
\begin{align}
    &A^{(p)}\w B^{(q)}=\frac{1}{(p+q)!}\left(\frac{(p+q)!}{p!q!}\,A_{[\m_1\m_2...\m_p}\,B_{\n_1\n_2...\n_q]}\right)\,dx^{\m_1}\w...dx^{\m_p}\w dx^{\n_1}\w...dx^{\n_q}\,. \\
    &A^{(p)}\w B^{(q)}\w...\w C^{(r)} =\frac{1}{(p+q+...r)!}\left(\frac{(p+q+...r)!}{p!q!...r!}\,A_{[\m_1...\m_p}B_{\n_1...\n_q}...C_{\r_1...\r_r]}\right)\nonumber\\
    &\qquad\qquad\qquad\qquad\qquad\qquad\quad\,dx^{\m_1}\w...dx^{\m_p}\w dx^{\n_1}\w...dx^{\n_q}...\w dx^{\r_1}\w...dx^{\r_r}\,.
\end{align}
\textbf{Interior product:}
\begin{align}
    &\i_v(A^{(p)})=\frac{1}{(p-1)!}\,\left(v^{\m_1}A_{[\m_1\m_2...\m_p]}\right)\,dx^{\m_2}\w dx^{\m_3}...\w dx^{\m_p}\,.
\end{align}
\textbf{Hodge dual:}
\beq
 \s (A^{(p)})=(\s A)^{(d-p)}=\frac{1}{(d-p)!}\left(\frac{1}{p!}\,\epsilon_{\m_1 \m_2...\m_d}\,A^{\m_1\m_2...\m_p}\right)\,dx^{\m_{(p+1)}}\w... d x^{\m_d}\,. 
\eeq
\textbf{Exterior derivative:}
\begin{align}
    &d(A^{(p)})=\frac{1}{(p+1)!}\left((p+1)\,\pa_{[\m_1}A_{\m_2...\m_{(p+1)}]}\right)\,dx^{\m_1}\w dx^{\m_2}\w...dx^{\m_{(p+1)}}\,.
\end{align}
\textbf{Integral}\\
On a $d$-dimensional manifold, for a form of the top rank $d$, the Hodge dual is a scalar. One can thus use the following definition for integrals of such forms via the ordinary invariant integral of a scalar:
\begin{align}
    \int C^{(d)}\equiv \int d^dx\, \frac{\sqrt{|g|}}{\,d!}\,\epsilon^{\m_1 \m_2 ...\m_d}C_{\m_1 \m_2 ...\m_d}.
\end{align}
Note that the explicit $\sqrt{|g|}$ is cancelled by the one contained in the Levi-Civita tensor as per (\ref{LCtens}), making this integration formula metric independent. It can be understood as simply $\int d^dx\,\, C_{12...d}$.

\section{Exterior calculus identities}\label{excalid}

The following identities are used extensively in the calculations of this work. Here, $u$ and $v$ are 1-forms (the same letters are used for the dual vectors), $A$ and $C$ are $p$-forms, and $B$ is a $q$-form. The number of dimensions is $d$. \vspace{2mm}\\ 
\textbf{Exterior derivative identities:}
\begin{align}
    dd&=0 \quad (\text{nilpotency}),\\
    d(A\w B)&=dA \w B +(-1)^p\,A\w dB \quad (\text{Leibniz rule}). \label{extderleibniz}
\end{align}
\textbf{Commutation in the exterior product:}
\beq
    A\w B = (-1)^{pq}\,B\w A.\label{commext}
\eeq
\textbf{Hodge star identities:}
\begin{align}
    \s \s A&=\text{sgn}(g)(-1)^{p(d-p)}A, \label{starstarid}\\
    A\w \s C&= C\w \s A \quad (\text{where both}\,\, A\,\, \text{and}\,\, C \,\,\text{have the same rank}), \label{AstarC}
\end{align}
where $\text{sgn}(g)=-1$ for Minkowski space (we use the mostly-plus metric signature).\vspace{2mm}\\
\textbf{Interior product identities:}
\begin{align}
    \i_u \i_v &=-\,\i_v \i_u,\qquad
    \i_v \i_v=0, \label{interiorprodnilpotency}\\
    \i_v(A\w B)&= \i_vA \w B +(-1)^p\,A\w \i_vB \quad (\text{Leibniz rule}) \label{interiorprodleibniz},\\
    \i_v \s A &= \s (A\w v), \label{ivstarid}\\
    \s (\i_v A)&=(-1)^{(p-1)}\, v\w \s A =(-1)^{(d-1)}\,\s A\w v, \label{starivid} \\
   \i_v(v\w A)+v\w \i_vA&=A \label{projrej} \quad \text{(projection-rejection decomposition)},
\end{align}
with $v^2=1$ assumed in the last line.

\section{General solution of the equation $da\w dB=0$}\label{appgensol}

We consider the general equation 
\begin{align}\label{dadB}
    da\w dB=0
\end{align}
where $B$ is a $p$-form in $d$ dimensions and $a$ is a scalar field that has the properties of a good global coordinate. In particular, the level surfaces of $a$ defined by equations $a=\text{const}$ provide a globally
nondegenerate foliation of $R^d$, and each slice $a=\text{const}$ has the topology $R^{d-1}$.

The key observation is that \eqref{dadB} has a natural and simple restriction to the level surfaces $a=\text{const}$. Then the easiest way to analyze \eqref{dadB} is to go to a coordinate system where one of the coordinates, which we shall call $x^0$, is simply chosen as $a$. We shall call the remaining coordinates, parametrizing the constant $a$ slices, $x^i$ with $i=1,\ldots, d-1$. In this coordinate system, \eqref{dadB} becomes
\begin{align}
\pa_{[i} B^\perp_{i_1\cdots i_p]}=0
\end{align}
where $B^\perp$ denotes the restriction of $B$ on the surface $a=\text{const}$ (a $p$-form in $R^{d-1}$) given by the components of $B$ along the surface $a=\text{const}$, $B^\perp_{i_1\cdots i_p}=B_{i_1\cdots i_p}$. The above equation simply says that $B^\perp$ is closed on the $a=\text{const}$ slice. Since the slice is isomorphic to $R^{d-1}$, this immediately implies that $B^\perp$ is exact, and the most general solution of \eqref{dadB} is 
\begin{align}
B^\perp_{i_1\cdots i_p}=\pa_{[i_1} C_{i_2\cdots i_p]},
\end{align}
where $C$ is an arbitrary $(p-1)$-form on each slice (which may also have an arbitrary parametric dependence on $x^0$, since the equations on different slices are completely independent). The solution above is equivalent to the following coordinate-invariant statement in the original Minkowski space:
\beq
da\w(B-dC)=0.
\eeq
For any form $A$, $da\w A=0$ implies that $A=da\w E$ for some form $E$. This is easily seen by first expressing $A$ using the projection-rejection identity (\ref{prre}) and then taking into account that $da\w A=0$ . Hence, the most general solution of the above equation, and of (\ref{dadB}), is
\beq
B=dC+da\w E,
\eeq
where $C$ and $E$ are arbitrary.

\section{PST-like form of the democratic Lagrangian}\label{appequiv}

\subsection{Derivation}

Lagrangian \eqref{polylag} can be rewritten as follows:
\begin{align}\label{equivLag}
    \mathcal{L}=F^2+G^2+(da\w R)^2 +(da\w S)^2 - 2\,\mathcal{F}\w \s (da\w R) +2\,da\w S\w\mathcal{F}
\end{align}
where $\mathcal{F}=(F-\s G)$. It has the following variation upto total derivatives:
\begin{align}
    \d \mathcal{L}=\,& -\,2\, \d a\,d[R\w \s(da\w R -\mathcal{F})+S\w\{\mathcal{F}+\s(da\w S)\}] \nonumber \\
    &\,-\,2\,(-1)^p\, \d A\w[d\s(F-da\w R)+(-1)^{(p+d+pd)}da\w dS] \nonumber \\
    &\, -\,2\,(-1)^{(d-p)}\, \d B\w [d\s(G+(-1)^{(p+d+pd)}da\w S)-da\w dR] \nonumber \\
    &\, +2\,(-1)^p\,\d R\w da\w\s(da\w R-\mathcal{F})+2\,(-1)^{(d-p)}\d S\w da\w[\mathcal{F}+\s(da\w S)]\,.
\end{align}
The equations of motion of the auxiliary fields $R$ and $S$ are
\begin{align}\label{eomRS}
    \frac{\d\mathcal{L}}{\d R}\equiv da\w\s(da\w R -\mathcal{F})=0\, \quad \text{and} \quad 
    \frac{\d\mathcal{L}}{\d S}\equiv da\w[\mathcal{F}+\s(da\w S)]=0\,. 
\end{align}
The above equations can be rewritten as following:
\begin{align}\label{rewriteomRS}
    R=\frac{1}{(\pa a)^2}\,(\i_{da} \mathcal{F}+da\w \i_{da}R)\,\quad \text{and} \quad
    S=-\,(-1)^{pd}\,\frac{1}{(\pa a)^2}\,(\i_{da} \mathcal{\s F}+da\w \i_{da}S)\,.
\end{align}
Due to the gauge freedom coming from symmetries \eqref{gaugeR} and \eqref{gaugeS}, the above equations imply that the following equations also hold:
\begin{align}
    R=\frac{1}{(\pa a)^2}\,(\i_{da} \mathcal{F}+da\w C)\,\quad \text{and} \quad
    S=-\,(-1)^{pd}\,\frac{1}{(\pa a)^2}\,(\i_{da} \mathcal{\s F}+da\w D)\,.
\end{align}
where $C$ is an arbitrary $(p-1)$-form and $D$ is arbitrary $(d-p-3)$-form. In the above solution for $R$ the second term can be gauged away using gauge symmetry \eqref{gaugeR} and the second term in the above solution for $S$ can be gauged away by transformation \eqref{gaugeS}. The we get the following solutions for $R$ and $S$:
\begin{align}\label{RSsol}
    R=\frac{\i_{da} \mathcal{F}}{(\partial a)^2}\,\quad \text{and} \quad
    S=-\,(-1)^{pd}\,\frac{\i_{da} \mathcal{\s F}}{(\partial a)^2}\,.
\end{align}
On plugging the above solutions into polynomial Lagrangian \eqref{equivLag} we get the democratic PST-like Lagrangian \eqref{demlag}:
\begin{align}\label{demlag}
    \mathcal{L}= F^2+G^2- (\i_v \mathcal{F})^2  - (\i_v \s \mathcal{F})^2
\end{align}
where $v={da}/\sqrt{({\p a})^2}$, $a$ is a scalar, 
$F=dA$, $A$ is a $p$-form gauge potential, $G=dB$, $B$ is a $(d-p-2)$-form gauge potential, and 
$\mathcal{F}\equiv F-\s G$.

The variation of Lagrangian \eqref{demlag} can be derived in the following form with methods typical of the PST theory (see, for instance, \cite{Hiroshi}), though this derivation is considerably more demanding than for the polynomial case:
\begin{align}\label{vardeml}
    \d\mathcal{L}&=(-1)^p\,4\,\d a\,\, d\bigg(\frac{v}{\sqrt{(\p a)^2}}\w \i_v\mathcal{F}\w \i_v\s \mathcal{F}\bigg) - (-1)^p\,4\,\d A\w d(v\w \i_v\s \mathcal{F}) \nonumber \\
    &\quad - (-1)^{(d-p)}\,4\,\d B\w d(v\w \i_v\mathcal{F}).
\end{align}
One has, in particular, to keep in mind the following variation of $v$:
\begin{align}\label{varv}
    \d v= \frac{\i_v(v\w d\d a)}{\sqrt{({\p a})^2}}\,.
\end{align}
Lagrangian \eqref{demlag} is invariant under the following gauge transformations:
\begin{align}
    \d a&=0\,, \quad \d A= dU\,, \quad \d B=0\,; \\
    \d a&=0\,, \quad \d A=0 , \quad \d B=dV\,\,; \\
    \d a&=0\,, \quad \d A= da\w U\,, \quad \d B=0\,\,; \label{demtrans2} \\
    \d a&=0\,, \quad \d A=0\,, \quad \d B=da\w V\,\,; \label{demtrans3} \\
    \d a&=\varphi\,, \quad \d A=\frac{\varphi}{\sqrt{(\p a)^2}}\,\i_v\mathcal{F}\,, \quad \d B=(-1)^{(p+d+pd)}\frac{\varphi}{\sqrt{(\p a)^2}}\,\i_v\s \mathcal{F}\,; \label{demgaugetrans}
\end{align}
where $\varphi$ is a $0$-form, $U$ is a $(p-1)$-form and $V$ is a $(d-p-3)$-form. The invariance can be verified using (\ref{vardeml}).

%%%%%%%%%%%%

\subsection{Dynamics}
The equations of motion are as follows:
\begin{align}
    \frac{\d \mathcal{L}}{\d a}&\equiv d\bigg(\frac{v}{\sqrt{(\p a)^2}}\w \i_v\mathcal{F}\w \i_v\s \mathcal{F}\bigg)=0, \label{aeom} \\
    \frac{\d \mathcal{L}}{\d A}&\equiv d(v\w \i_v\s \mathcal{F})=0, \label{Aeom} \\
    \frac{\d \mathcal{L}}{\d B}&\equiv d(v\w \i_v\mathcal{F})=0. \label{Beom}
\end{align}
When equations \eqref{Aeom} and \eqref{Beom} hold, \eqref{aeom} holds identically. This can be seen using the relations
\beq 
d\bigg(\frac{1}{\sqrt{(\p a)^2}}\bigg)\w v = \frac{dv}{\sqrt{(\p a)^2}}\,,
\qquad d\bigg(\frac{v}{\sqrt{(\p a)^2}} \bigg)=2\, \frac{dv}{\sqrt{(\p a)^2}}\,.
\eeq 

One can rewrite \eqref{Beom} as
\begin{align}\label{Beomrewrit}
    da\w d\bigg(\frac{\i_v\mathcal{F}}{{\sqrt{(\p a)^2}}}\bigg)=0\,.
\end{align}
As per analysis of appendix~\ref{appgensol}, the general solution to \eqref{Beom} is
\begin{align}
    v\w \i_v\mathcal{F}&=-\,da\w d X,
\end{align}
where $X$ is an arbitrary $(p-1)$-form.
Under the gauge transformation \eqref{demtrans2} $v\w \i_v\mathcal{F}$ transforms as
\begin{align}
    v\w \i_v\mathcal{F}\rightarrow v\w \i_v\mathcal{F}-da\w dU=-\,da\w d X-da\w dU.
\end{align}
Fixing the value of the gauge field $U$ as $U=-X$ we get
\begin{align}\label{vivF0}
    v\w \i_v\mathcal{F}=0\,.
\end{align}
The general solution to \eqref{Aeom} is
\begin{align}
    v\w \i_v\s\mathcal{F}&=d(da\w Y)=-\,da\w d Y.
\end{align}
where $Y$ is an arbitrary $(d-p-2)$-form.
Under the gauge transformation \eqref{demtrans3} $v\w \i_v\s\mathcal{F}$ transforms as
\begin{align}
    v\w \i_v\s\mathcal{F}\rightarrow v\w \i_v\s\mathcal{F}+(-1)^{(p+pd)}\,da\w dV=-\,da\w d Y+(-1)^{(p+pd)}\,da\w dV.
\end{align}
Fixing the value of the gauge field $V$ as $V=(-1)^{(p+pd)}\,Y$ we get
\begin{align}\label{vivsF0}
    v\w \i_v\s\mathcal{F}=0\,.
\end{align}
Using the projection-rejection identity (\ref{prre}), as well as (\ref{starivv}), (\ref{vivF0}) and (\ref{vivsF0}) imply $\mathcal{F}=0$, and hence
\beq
 F=\s G.
\eeq
This shows that, for all solutions of the equations of motion, the forms $F$ and $G$ are dual to each other, up to gauge redundancy.

When we take $d=2p+2$, $p$ to be even and $G=F$, the Lagrangian in \eqref{demlag} becomes
\begin{align}\label{usualPST}
    \mathcal{L}= 2\,[F^2- (\i_v \mathcal{F})^2]\,.
\end{align}
This is twice the usual PST Lagrangian \cite{PST3}.


\begin{thebibliography}{99}

\bibitem{Zw}D.~Zwanziger, {\it Local Lagrangian quantum field theory of electric and magnetic charges,} Phys.\ Rev.\ D \textbf{3} (1971) 880.

\bibitem{DT}S.~Deser and C.~Teitelboim, {\it Duality transformations of Abelian and non-Abelian gauge fields,} Phys.\ Rev.\ D \textbf{13} (1976) 1592.

\bibitem{MSch}N.~Marcus and J.~H.~Schwarz, {\it Field theories that have no manifestly Lorentz-invariant formulation,} Phys.\ Lett.\ B {\bf 115} (1982) 111.

\bibitem{Sieg}W.\ Siegel, {\it Manifest Lorentz invariance sometimes requires nonlinearity,} Nucl.\ Phys.\ B {\bf 238} (1984) 307–316.

\bibitem{KMkrtch}A.~R.~Kavalov and R.~L.~Mkrtchian, {\it Lagrangian of the selfduality equation and d=10, N=2b supergravity,} Sov.\ J.\ Nucl.\ Phys. {\bf 46} (1987) 728.

\bibitem{FJ}R.~Floreanini and R.~Jackiw, {\it Selfdual fields as charge density solitons,} Phys.\ Rev.\ Lett. {\bf 59} (1987) 1873.

\bibitem{HT}M.~Henneaux and C.~Teitelboim, {\it Dynamics of chiral (selfdual) p-forms,} Phys. Lett. B {\bf 206} (1988) 650.

\bibitem{BS}
M.~Blagojevi\'c and P.~Senjanovi\'c,
{\it The quantum field theory of electric and magnetic charge,}
Phys. Rept. \textbf{157} (1988) 233.

\bibitem{Harada}K.~Harada, {\it The chiral Schwinger model in terms of chiral bosonization,} Phys.\ Rev.\ Lett. {\bf 64} (1990) 139.

\bibitem{Tse1}A.~A.~Tseytlin, {\it Duality symmetric formulation of string world sheet dynamics,} Phys.\ Lett.\ B {\bf 242} (1990) 163.

\bibitem{MYW} B.~McClain, F.~Yu and Y.~S.~Wu, {\it Covariant quantization of chiral bosons and $OSp(1,1|2)$ symmetry,} Nucl.\ Phys.\ B {\bf 343} (1990) 689.

\bibitem{Wot}C.~Wotzasek, {\it The Wess-Zumino term for chiral bosons,} Phys.\ Rev.\ Lett. {\bf 66} (1991) 129.

\bibitem{Tse2}A.~A.~Tseytlin, {\it Duality symmetric closed string theory and interacting chiral scalars,} Nucl. Phys. B {\bf 350} (1991) 395.

\bibitem{SchS}
J.~H.~Schwarz and A.~Sen,
{\it Duality symmetric actions,}
Nucl. Phys. B \textbf{411} (1994) 35
\arXiv{hep-th/9304154}.

\bibitem{KhP}
A.~Khoudeir and N.~Pantoja,
{\it Covariant duality symmetric actions,}
Phys. Rev. D \textbf{53} (1996) 5974
\arXiv{hep-th/9411235}.

\bibitem{PST1}
P.~Pasti, D.~P.~Sorokin and M.~Tonin,
{\it Note on manifest Lorentz and general coordinate invariance in duality symmetric models,}
Phys. Lett. B \textbf{352} (1995) 59
\arXiv{hep-th/9503182} [hep-th].

\bibitem{PST2}
P.~Pasti, D.~P.~Sorokin and M.~Tonin,
{\it Duality symmetric actions with manifest space-time symmetries,}
Phys. Rev. D \textbf{52} (1995) 4277
\arXiv{hep-th/9506109}.

\bibitem{PSTproc}P.~Pasti, D.~P.~Sorokin and M.~Tonin, {\it Space-time symmetries in duality symmetric models,} in
{\it Gauge theories, applied supersymmetry, quantum gravity} (Leuven, 1995) pp. 167–176 \arXiv{hep-th/9509052}.

\bibitem{Tse3}
A.~A.~Tseytlin,
{\it Selfduality of Born-Infeld action and Dirichlet three-brane of type IIB superstring theory,}
Nucl. Phys. B \textbf{469} (1996) 51
\arXiv{hep-th/9602064}.

\bibitem{DH}
F.~P.~Devecchi and M.~Henneaux,
{\it Covariant path integral for chiral p-forms,}
Phys. Rev. D \textbf{54} (1996) 1606
\arXiv{hep-th/9603031} [hep-th].

\bibitem{PSch}
M.~Perry and J.~H.~Schwarz,
{\it Interacting chiral gauge fields in six-dimensions and Born-Infeld theory,}
Nucl. Phys. B \textbf{489} (1997) 47
\arXiv{hep-th/9611065}.

\bibitem{PST3}
P.~Pasti, D.~P.~Sorokin and M.~Tonin,
{\it On Lorentz invariant actions for chiral p-forms,}
Phys. Rev. D \textbf{55} (1997) 6292
\arXiv{hep-th/9611100}.

\bibitem{S5M}
I.~A.~Bandos, K.~Lechner, A.~Nurmagambetov, P.~Pasti, D.~P.~Sorokin and M.~Tonin,
{\it Covariant action for the superfive-brane of M-theory,}
Phys. Rev. Lett. \textbf{78} (1997) 4332
\arXiv{hep-th/9701149}.

\bibitem{CW}
M.~Cederwall and A.~Westerberg,
{\it Worldvolume fields, SL(2,Z) and duality: the type IIB three-brane,}
JHEP \textbf{02} (1998) 004
\arXiv{hep-th/9710007}.

\bibitem{BBS}
I.~A.~Bandos, N.~Berkovits and D.~P.~Sorokin,
{\it Duality symmetric eleven-dimensional supergravity and its coupling to M-branes,}
Nucl. Phys. B \textbf{522} (1998) 214
\arXiv{hep-th/9711055}.

\bibitem{MPS}
A.~Maznytsia, C.~R.~Preitschopf and D.~P.~Sorokin,
{\it Duality of selfdual actions,}
Nucl. Phys. B \textbf{539} (1999) 438
\arXiv{hep-th/9805110}.

\bibitem{PSTnotoph}
P.~Pasti, D.~P.~Sorokin and M.~Tonin,
{\it Harmonics, notophs and chiral bosons,}
Lect. Notes Phys. \textbf{524} (1999) 97
\arXiv{hep-th/9807133}.

\bibitem{RTse}
M.~Ro\v{c}ek and A.~A.~Tseytlin,
{\it Partial breaking of global D = 4 supersymmetry, constrained superfields, and three-brane actions,}
Phys. Rev. D \textbf{59} (1999) 106001
\arXiv{hep-th/9811232} [hep-th].

\bibitem{MMMK}
R.~Manvelyan, R.~Mkrtchian and H.~J.~W.~Muller-Kirsten,
{\it On different formulations of chiral bosons,}
Phys. Lett. B \textbf{453} (1999) 258
\arXiv{hep-th/9901084}.

\bibitem{LM}
K.~Lechner and P.~A.~Marchetti,
{\it Duality invariant quantum field theories of charges and monopoles,}
Nucl. Phys. B \textbf{569} (2000) 529
\arXiv{hep-th/9906079} [hep-th].

\bibitem{KT}
S.~M.~Kuzenko and S.~Theisen,
{\it Supersymmetric duality rotations,}
JHEP \textbf{03} (2000) 034
\arXiv{hep-th/0001068} [hep-th].

\bibitem{MMMK2}
Y.~G.~Miao, R.~Manvelyan and H.~J.~W.~Mueller-Kirsten,
{\it Selfduality beyond chiral p-form actions,}
Phys. Lett. B \textbf{482} (2000) 264
\arXiv{hep-th/0002060} [hep-th].

\bibitem{Sorokin}D.~Sorokin, {\it Lagrangian description of duality-symmetric fields,} NATO\ Sci.\ Ser.\ II {\bf 60} (2002) 365.



\bibitem{PST4}
P.~Pasti, D.~Sorokin and M.~Tonin,
{\it Covariant actions for models with non-linear twisted self-duality,}
Phys. Rev. D \textbf{86} (2012) 045013
\arXiv{1205.4243} [hep-th].

\bibitem{BH}
C.~Bunster and M.~Henneaux,
{\it Duality invariance implies Poincar\'e invariance,}
Phys. Rev. Lett. \textbf{110} (2013) 011603
\arXiv{1208.6302} [hep-th].

\bibitem{KSV}
S.-L.~Ko, D.~Sorokin and P.~Vanichchapongjaroen,
{\it The M5-brane action revisited,}
JHEP \textbf{11} (2013) 072
\arXiv{1308.2231} [hep-th].

\bibitem{Hiroshi}H.~Isono,
{\it Note on the self-duality of gauge fields in topologically nontrivial spacetime,}
PTEP \textbf{2014} (2014) 093B05
\arXiv{1406.6023} [hep-th].

\bibitem{Sen1}
A.~Sen, {\it Covariant action for type IIB supergravity,}
JHEP \textbf{07} (2016) 017
\arXiv{1511.08220} [hep-th].

\bibitem{AEShJ}
H.~Afshar, E.~Esmaeili and M.~M.~Sheikh-Jabbari,
{\it Asymptotic symmetries in $p$-form theories,}
JHEP \textbf{05} (2018) 042
\arXiv{1801.07752} [hep-th].

\bibitem{Sen2}
A.~Sen,
{\it Self-dual forms: action, Hamiltonian and compactification,}
J. Phys. A \textbf{53} (2020) 084002
\arXiv{1903.12196} [hep-th].

\bibitem{Buratti:2019cbm}
G.~Buratti, K.~Lechner and L.~Melotti,
{\it Duality invariant self-interactions of abelian p-forms in arbitrary dimensions,}
JHEP \textbf{09} (2019) 022
\arXiv{1906.07094} [hep-th].

\bibitem{Buratti:2019guq}
G.~Buratti, K.~Lechner and L.~Melotti,
{\it Self-interacting chiral p-forms in higher dimensions,}
Phys. Lett. B \textbf{798} (2019) 135018
\arXiv{1909.10404} [hep-th].

\bibitem{Kiral}
K.~Mkrtchyan,
{\it On covariant actions for chiral $p$-forms,}
JHEP \textbf{12} (2019) 076
\arXiv{1908.01789} [hep-th].

\bibitem{Lambert:2019diy}
N.~Lambert,
{\it (2,0) Lagrangian structures,}
Phys. Lett. B \textbf{798} (2019) 134948
\arXiv{1908.10752} [hep-th].

\bibitem{Townsend1}
P.~K.~Townsend,
{\it An interacting conformal chiral 2-form electrodynamics in six dimensions,}
Proc. Roy. Soc. Lond. A \textbf{476} (2020) 20190863
\arXiv{1911.01161} [hep-th].

\bibitem{Townsend2}
P.~K.~Townsend,
{\it Manifestly Lorentz invariant chiral boson action,}
Phys. Rev. Lett. \textbf{124} (2020) 101604
\arXiv{1912.04773} [hep-th].

\bibitem{Andriolo:2020ykk}
E.~Andriolo, N.~Lambert and C.~Papageorgakis,
{\it Geometrical aspects of an Abelian (2,0) action,}
JHEP \textbf{04} (2020) 200
\arXiv{2003.10567} [hep-th].

\bibitem{BLST}
I.~Bandos, K.~Lechner, D.~Sorokin and P.~K.~Townsend,
{\it A non-linear duality-invariant conformal extension of Maxwell's equations,}
Phys. Rev. D \textbf{102} (2020) 121703
\arXiv{2007.09092} [hep-th].

\bibitem{Bertrand:2020nob}
Y.~Bertrand, S.~Hohenegger, O.~Hohm and H.~Samtleben,
{\it Toward exotic 6D supergravities,}
\arXiv{2007.11644} [hep-th].

\bibitem{Kosyakov}
B.~P.~Kosyakov,
{\it Nonlinear electrodynamics with the maximum allowable symmetries,}
Phys. Lett. B \textbf{810} (2020) 135840
\arXiv{2007.13878} [hep-th].

\bibitem{Vanichchapongjaroen}
P.~Vanichchapongjaroen,
{\it Covariant M5-brane action with self-dual 3-form,}
\arXiv{2011.14384} [hep-th].

\bibitem{Bandos:2020hgy}
I.~Bandos, K.~Lechner, D.~Sorokin and P.~K.~Townsend,
{\it On p-form gauge theories and their conformal limits,}
\arXiv{2012.09286} [hep-th].

\bibitem{Cremonini:2020skt}
C.~A.~Cremonini and P.~A.~Grassi,
{\it Self-dual forms in supergeometry I: the chiral boson,}
\arXiv{2012.10243} [hep-th].

\bibitem{BB1}I.~Bia\lslash ynicki-Birula, {\it Nonlinear electrodynamics: variations on a theme by Born and Infeld,} in {\it Quantum theory of particles and fields: birthday volume dedicated to Jan \Lslash opusza\'nski} (World Scientific, 1984), pp. 31-48.

\bibitem{BB}
I.~Bia\lslash ynicki-Birula,
{\it Field theory of photon dust,}
Acta Phys. Polon. B \textbf{23} (1992) 553.

\bibitem{Chruscinski}
D.~Chru\'sci\'nski,
{\it Strong field limit of the Born-Infeld p-form electrodynamics,}
Phys. Rev. D \textbf{62} (2000) 105007
\arXiv{hep-th/0005215}.

\end{thebibliography}
\end{document}